\documentstyle[preprint,aps]{revtex}
\begin{document}
\draft
\preprint{}

\title{Neutron Drops and Skyrme Energy-Density Functionals}
\author{B. S. Pudliner$^{a}$, A. Smerzi$^{a,b}$, J. Carlson$^{c}$,\\ V. R.
Pandharipande$^{a}$, Steven C. Pieper$^{d}$, D. G. Ravenhall$^{a}$}
\address{a. Department of Physics, University of Illinois, Urbana,
IL 61801-3080}
\address{b. Laboratorio Nazionale del Sud, INFN, v Andrea Doria,
95125 Catania, Italy}
\address{c. Theoretical Division, Los Alamos National Laboratory,
Los Alamos New Mexico 87545}
\address{d. Physics Division, Argonne National Laboratory,
Argonne, IL 60439-4843}

\date{\today}

\maketitle


\begin{abstract}
The J$^{\pi}$=0$^+$ ground state of a drop of 8 neutrons and the lowest
1/2$^-$ and 3/2$^-$ states of 7-neutron drops, all in an external well,
are computed accurately
with variational and  Green's function Monte Carlo methods for a Hamiltonian
containing the Argonne $v_{18}$ two-nucleon and Urbana IX three-nucleon
potentials.
These states are also calculated using Skyrme-type energy-density functionals.
Commonly used functionals overestimate the central density of these
drops and the spin-orbit splitting of 7-neutron drops.  Improvements in the
functionals are suggested.
\end{abstract}
\vspace{.5in}

Properties of neutron matter are vitally important in determining the
structure of neutron stars \cite{refa}, and have a strong
bearing on the energies of neutron-rich nuclei,
and on the r-process in  nucleosynthesis \cite{refb}.
It is impossible to extrapolate available data on nuclei to the
region of neutron matter with sufficient precision
using effective interactions.
Different effective interactions that fit the energies of laboratory nuclei
rather well predict very different equations of state for
neutron matter \cite{refa}.
In contrast, it appears that modern calculations of neutron matter
based on realistic models of
nuclear forces are much more consistent with each other at
densities $\rho \alt 0.16 fm^{-3}$\cite{prrev}, and therefore are
presumably more reliable.
The two-nucleon interaction in these realistic models is better
determined from the scattering data in isospin T=1 states than
that in T=0, and the
uncertainties coming from three-nucleon forces and relativistic
effects are also much smaller in neutron than in nuclear matter.
Calculations of uniform neutron matter
have provided important constraints on Skyrme-type
effective interactions used to study neutron-rich systems.
They do not, however, provide information on the strength of the spin-orbit
interaction, nor on other terms sensitive to density gradients,
both of which may affect significantly the predicted properties
of drip-line nuclei and of neutron-star matter.

Ab initio calculations of finite nuclei, based on realistic
models of nuclear forces, can provide the necessary additional
information, but they are more challenging.
Recently \cite{pudliner95} the energies of nuclei with A $\leq$ 6
have been calculated essentially exactly with
the Green's function Monte Carlo (GFMC) method.
Cluster variational Monte Carlo (CVMC) calculations have also been used
to study $^{16}$O \cite{pieper92} and the spin-orbit
splitting (SOS) in $^{15}$N \cite{pieper93}.
In this letter we report  GFMC and CVMC calculations of states of
seven and eight neutrons bound in a weak external potential well using
the new Argonne two-nucleon \cite{wiringa95} and Urbana
three-nucleon interactions used in Ref.\ \cite{pudliner95}.
These interactions accurately reproduce the available two-nucleon
scattering data and binding energies of A $\leq$ 6 nuclei.
Neutron matter is not bound, therefore an
external well ($V_{ex}$) is necessary to hold the neutrons together.
We have used a Woods-Saxon well
with $V_o = -20$ MeV, $R = 3$ fm, and $a = 0.65$ fm,
chosen such that with it alone only the single-neutron 1s
state is bound at -5.73 MeV, while the 1p and higher states are unbound.
The investigated states of seven and eight neutrons are thus bound by
both the well and the interaction between neutrons.
We denote them by $^8$n$(J^{\pi} = 0^{+})$ and $^7$n$(J^{\pi}
= 1/2^{-}$ and 3/2$^-$).

The present variational Monte Carlo (VMC) and GFMC
calculations are simpler than those for
nuclei \cite{pudliner95} because all nucleons are neutrons.
The wave function is represented by a vector function of
$\vec{R}$ $(\equiv \vec{r}_1, \vec{r}_2, \cdots, \vec{r}_{A} )$
with 2$^A$ spin components specifying the spin direction of each neutron.
The VMC and GFMC calculations use a simpler
variational wave function than those of Refs \cite{pudliner95,pieper92}:
\begin{equation}
|\Psi_{V}\rangle = \left[ S \, \prod_{i<j}  (1 + U_{ij} )\right]
\left[ \prod_{i<j}\, f_c (r_{ij} ) \right] |\Phi\rangle \,,
\label{eq:f6}
\end{equation}

\begin{equation}
U_{ij} = u_{\sigma} (r_{ij} )\, \sigma_i \cdot \sigma_j + u_t (r_{ij} ) S_{ij}
\,.
\end{equation}
Here $S\Pi$ denotes a symmetrized product, $S_{ij}$ is the tensor operator,
$f_c (r_{ij} )$ is the Jastrow correlation,
and $|\Phi\rangle$ is an antisymmetric shell model wave function.
The three-body correlations commonly used in nuclear $\Psi_{V}$
are omitted because they have little effect on the energies of low-density
neutron systems,
and the two-body spin-orbit correlations
are discussed later along with an improved $\Psi_{V}$.
The radial wave functions of the
s- and p-orbitals in $\Phi$ and the correlation functions
$f_c$, $u_{\sigma}$ and $u_t$ are determined variationally.

The GFMC calculations are carried out as described in Ref.\ \cite{pudliner95}
with a simpler Hamiltonian:
\begin{equation}
H = -\sum_{i}{\hbar^2\over 2m} \nabla^{2}_{i} + \sum_{i} V_{ex}(i)
 + \sum_{i<j} v^{'}_{8} (ij) + \sum_{i<j<k} V_{ijk} \,,
\end{equation}
where the $v^{'}_{8}$ does not contain $L^2$  or $(L\cdot S)^2$ terms;
it equals the charge-symmetric part
of the Argonne $v_{18}$ interaction \cite{wiringa95} in the $^1S_0$ and
$^3P_{J=0,1,2}$ two-neutron states.
The small difference between the full $v_{18}$ and $v^{'}_{8}$
is treated as a first-order perturbation, whose contribution to the
calculated energies is $<$ 0.2 MeV.

The calculated transient energies, $E(\tau)$,
\begin{equation}
E(\tau ) = \langle \Psi_{V} |H e^{-(H-E_{o})\tau}| \Psi_{V} \rangle /
\langle \Psi _{V} |e^{-(H-E_{o})\tau}| \Psi_{V} \rangle\,,
\end{equation}
are shown in Fig.1.  The $E(\tau \rightarrow \infty)$ converges to the
lowest eigenvalue of the chosen $J^{\pi}$.  In Fermi systems, the statistical
error in $E(\tau)$ increases with $\tau$ as configurations diffuse
across nodal surfaces of the wave function\cite{signprob}.
Due to the large number of nodal surfaces in the wave functions of $^7$n
and $^8$n states, it is difficult to study their transient energy for
values of $\tau > 0.04$ MeV$^{-1}$.
The average values of
$E (\tau )$ for $\tau$ = 0.032, 0.036 and 0.04 MeV$^{-1}$, denoted
by $\bar E$, are shown by horizontal lines in Fig.\ 1.
The $E(\tau )$ of the $^8$n$(0^+ )$ and  $^7$n(1/2$^-$)
states do not have much $\tau$-dependence for $\tau >$ 0.015 MeV$^{-1}$,
suggesting that their $\bar E$ can be identified with the eigenvalues.
In contrast, the $E(\tau )$
of the $^7$n$(3/2 ^-$) state has more $\tau$-dependence.
Consequently, the eigenvalue of the lowest $^7$n$(3/2^- )$ state
could be a little below its $\bar E$ value. However,
we will neglect that difference
and regard it as our best estimate of the eigenvalue.
The GFMC estimate of the density distribution of neutrons in the  $^8$n$(0^+ )$
drop is
shown in Fig. 2. These results can be used to test the accuracy of the CVMC
method and to further constrain the Skyrme type energy-density functionals
used to study neutron-rich nuclei and neutron star crusts as discussed below.

The CVMC method and its modification for SOS are described in
Refs.\ \cite{pieper92,pieper93}.  The present CVMC
calculations are more accurate; they include contributions of all correlations
and interactions up to five-body clusters.
In contrast, in \cite{pieper92,pieper93} contributions of only static
correlations and interactions were calculated up to four-body clusters and
the momentum-dependent terms were evaluated only at the two-body level.
With the simpler $\Psi_{V}$ given by Eq.\ (\ref{eq:f6}), the 1- to 5-body
cluster contributions to the energy of $^8$n$(0^+ )$ state
are respectively 12.9, -54.5, 11.1, -3.8, and 1.1 MeV,
which sum up to -33.3(2) MeV.   The $E(\tau = 0)$ is nothing but the
variational energy calculated to all orders without cluster expansion.
Its value of -33.7(1) MeV is very close to the CVMC result retaining
up to five-body clusters.  Note that even in this rather low-density
system, the cluster expansion has a slow convergence and it
appears necessary to include five-body cluster contributions to reduce
the truncation error to  $<$ 2\%.

	The simple $\Psi_{V}$ (Eq.\ \ref{eq:f6}) is not very accurate;
the energy obtained with it is $\sim$ 4 MeV (or $\sim$ 11\%) too large.
In CVMC we use the more general variational wave functions of the form:
\begin{equation}
|\Psi^{'}_{V} \rangle = \left[ 1 + \sum_{i<j<k} U_{ijk} \right]
\left[ S\prod_{i<j} (1 + U_{ij} ) \right] \left[ 1 + \sum_{i<j}
u_{L\cdot S}  (r_{ij}) L_{ij} \cdot (\sigma _i + \sigma _ j )\right]
\left[ \prod_{i<j} f_c (r_{ij} ) \right] |\Phi \rangle \label{eq:f8}
\end{equation}
where the three-body correlations, $U_{ijk}$, are of the
kind used in Refs.\ \cite{pieper92,pieper93}
(note, however, that the commutator term is zero in pure neutron systems),
and as before the $U_{ij}$ contains spin and tensor terms.  The energies
obtained with the $\Psi^{'}_{V}$ variational wave functions are
respectively -35.6(1), -31.2(1), -29.7(1) MeV for the $^{8}$n(0$^+$),
$^{7}$n($\frac{1}{2}^{-}$) and $^7$n($\frac{3}{2}^{-}$) states.
They are only $\sim$ 4 \%
above the GFMC energies -37.6(3), -32.3(2) and -31.2(2)  MeV.
The CVMC calculations require about a factor of 25 less computer time
than the GFMC, even allowing for the variational search.  It is not difficult
to reduce the statistical error in CVMC calculations to a
fraction of one percent.
Much of the improvement in $\Psi_{V}^{\prime}$ comes from the
spin-orbit correlations omitted in the simpler $\Psi_{V}$.
In the present GFMC calculations, the spin-orbit correlations are built in
exactly via the propagation in imaginary time.

Fragmentation of the p$_{3/2}$ strength in $^{15}$N results in the SOS
of p$_{1/2}$ and p$_{3/2}$ quasi-hole states
being $\sim$0.6 MeV larger than the observed splitting between the lowest
3/2$^-$ and 1/2$^-$ states in $^{15}$N \cite{pieper93,n15frag}.
If the 3/2$^-$ hole strength in $^7$n is similarly fragmented,
the difference in the energies of lowest $^7$n(1/2$^-$)
and $^7$n$(3/2 ^-$) states could be smaller
than the SOS in $^7$n.  If $W$ denotes the energy width of
the fragmentation, the
GFMC transient energies, $E(\tau )$, for $\tau \ge 1/W$
will include fragmentation effects.  Since W is expected to be only a few MeV,
it is very unlikely that the present $\bar E$ evaluated up
to $\tau \sim 0.04$ has any fragmentation effects.
The $^7$n variational wave functions are constructed by removing an
appropriate state from the $|\Phi\rangle$ and thus correspond to
quasi-hole states.  Presumably, they too do not contain any fragmentation
effects.  Hence, we identify the difference between the calculated energies
of the $^7$n(1/2$^-$) and (3/2$^-$) states as the spin-orbit splitting.  Its
value is 1.1$\pm$0.3 and 1.4$\pm$0.1 MeV in the GFMC and
CVMC calculations respectively.

The energies of uniform neutron matter calculated from several
realistic models of nuclear forces were plotted in Ref.\ \cite{refa}
and compared with the results of four energy-density
functionals (EDF) used for astrophysical investigations involving
dense matter:
(i) Skyrme~1$^{\prime}$ - Vautherin-Brink Skyrme model 1 \cite{vb} modified
\cite{rbp} to fit the neutron-matter $E(\rho)$ of Ref.\ \cite{sp};
(ii) SkM - Skyrme model $M$ \cite{skm};
(iii) FPS - a generalized Skyrme model fitted approximately \cite{fps1,refa}
to the nuclear- and neutron-matter energies of Ref.\ \cite{fp};
and (iv) FPS21 - a generalized Skyrme model \cite{refa} fitted accurately to
results of Ref.\ \cite{fp}.
These EDF's reproduce the ground-state energies of stable closed-shell nuclei
rather accurately.  The root mean square deviations $|\Delta E/E|$ between
their prediction and experiment for
$^{16}$O, $^{40}$Ca, $^{48}$Ca, $^{56}$Ni, $^{90}$Zr, $^{114}$Sn, $^{140}$Ce,
and $^{208}$Pb are listed in Table~I.
This table also contains their predictions for the energy of the
$^8$n ground state and the $^7$n and $^{15}$N SOS.
The results for the density distributions $\rho (r)$ of the
$^8$n$(0^+ )$ state are compared with the GFMC and CVMC $\rho$(r) in Fig.\ 2.

In trying to learn from the departures of the EDF results for $^8$n and $^7$n
from our benchmarks, we concentrate on the FPS21 effective interaction,
since it gives the closest fit to the neutron-matter energies.
Possible sources of difference include the fact that the neutron-matter
energies used\cite{fp} date from an earlier period,
whereas the benchmark results
use newer interactions and more subtle computational techniques;
also, and more importantly, the density-gradient terms in FPS21 are related
to the effective-mass results\cite{fp} assuming a zero-range
nucleon-nucleon interaction with no spin exchange.  The latter simplification
is common to all of the EDF's we consider and is not well justified.
On the assumption that this is a contributor to the discrepancy, we have
examined the effect of adding a term,
$\frac{1}{2} \alpha  (\rho_n)^{\beta} (\nabla \rho_n)^2$, to the
FPS21 neutron EDF.
The gradient term,
$\frac{1}{2} \alpha (\rho_n + \rho_p)^{\beta} (\nabla \rho_n - \nabla
\rho_p)^2$,
reduces to the form used for
neutron drops, and would give little contribution for $N \sim Z$ nuclei.
However, in this work we have used it only for the neutron drops
to avoid refitting the models to laboratory nuclei.
Such a term can correct for
the overbinding of $^8$n$(0^+)$, and also reduce the central neutron densities.
We find that it cannot give good fits to these
quantities simultaneously, however.
For the two sets of coefficients (${\beta}, \alpha) = (0, 150$ MeV fm$^8$) and
($2, 7\times10^4$MeV fm$^{14}$) the ground state energies of $^8$n
are $-39.6$MeV and $-40.2$MeV respectively, and the neutron density
distributions are shown as curves A and B in Fig.\ 2.

The present CVMC and GFMC results clearly indicate that the SOS
predicted by the unadjusted Skyrme models for $^7$n is too large,
while it is good for $^{15}$N.
Relativistic mean-field models also predict a weaker spin-orbit potential
in neutron-rich nuclei  \cite{dhns}.
The neutron spin-orbit potential in these Skyrme models is of the form
\begin{equation}
V^{(n)}_{\ell s} (r) = W_{\ell s} {1\over r}\, {d\over dr}\, (\rho \,(r) +
\rho_n (r) ) \,, \label{eq:vb}
\end{equation}
obtained by Vautherin and Brink\cite{vb} assuming that it originates
from two-nucleon interactions in the triplet-P state.
Here $\rho (r) = \rho _n (r) + \rho _p (r)$ and
$W_{\ell s}$ is a constant determined from the SOS in laboratory nuclei like
$^{15}$N, and listed in Table~I.  The form of this potential,
proportional to a radial derivative
of the densities, indicates that apart from other dependences,
the SOS obtained with a given EDF will depend on the
nucleon central densities given by that model.
The two modified versions of FPS21 just described, for the
same value $W_{\ell s} = 110$MeV used in Table~I, each give a SOS in
$^7$n of 2.2 MeV.
The reduction from the value of 3 MeV for unmodified FPS21
occurs because of the reduced central densities induced by
the extra gradient term. The SOS in these modified models is still about
double the value predicted by CVMC and GFMC, however.

As discussed in Ref.\ \cite{pieper93}, more than
half the SOS in $^{15} $N comes from
three-nucleon contributions involving either two neutrons and a proton or
vice-versa.   In contrast, the three-body interaction
and clusters give a very small contribution to the SOS in $^7$n in CVMC
calculations.
This suggests adding terms to the EDF that will produce
a spin-orbit potential:
\begin{equation}
V^{(n)}_{\ell s} (r) = W^{(2)}_{\ell s} {1\over r}\, {d\over dr} (\rho \,(r) +
\rho _n (r)) +
W^{(3)}_{1,\ell s} {1\over r}\, {d\over dr} (\rho _n \, (r) \rho _p (r)) +
W^{(3)}_{2,\ell s} {1\over r}\, {d\over dr} (\rho _p (r) )^{2}
\label{eq:vbnew}
\end{equation}
having separate two- and three-body contributions.
The $(\rho _n (r) )^{2}$ term is omitted because three-neutron clusters seem
to give negligible $V^{(n)}_{\ell s} (r)$.
In neutron drops only the two-body $W^{(2)}_{\ell s}$ contributes, while in
N $\sim$ Z nuclei, like $^{15} $N,
$\rho_{n}(r)\rho_{p}(r) \sim \rho_{p}(r)^{2}$ and the sum
$W^{(3)}_{\ell s} = W^{(3)}_{1,\ell s} + W^{(3)}_{2,\ell s}$
is the only relevant new parameter.

With this modification to the spin-orbit interaction,
and the FPS21 parameterization for the central interaction (including the
gradient term for the neutron drops but not for $^{15}$N), we
can fit the spin-orbit splittings of $^{15}$N and $^7$n exactly,
using the parameter values
$W^{(2)}_{\ell s}$ = 61 MeV~fm$^5$ and $W^{(3)}_{\ell s}$ = 745 MeV~fm$^8$.
The spin-orbit splittings in the eight closed-shell nuclei mentioned earlier
are
modified only slightly, a not unexpected result in view of
their relatively small values of neutron excess

In conclusion, we have made the first exact microscopic calculations of
neutron drops in an external potential well.
Our results suggest that the commonly
used EDF's need modification in order
to describe accurately neutron-rich nuclei.  It appears
that they predict neutron drops which are too dense and have too large
a spin-orbit splitting.  Additional density gradient terms
need to be considered and the parameterization
of the Skyrme $V_{\ell s}$ must be modified to include three-body
contributions.
Our GFMC results also show that the CVMC using the
improved $\Psi^{\prime}_{V}$, with two-neutron spin-orbit correlations
gives fairly accurate results.
We plan to use CVMC to calculate the properties of larger neutron drops.
This, together with data from stable nuclei, will provide a larger
database for fitting a Skyrme EDF for studies of stable nuclei,
neutron rich nuclei, and the surface of neutron stars.

\section*{Acknowledgments}
We thank Drs. J. Dobaczewski and C. Gaarde for very useful advice.
These calculations were made possible by generous grants of time on the
IBM SP computer at the Mathematics and Computer Science Division, Argonne
National Laboratory.
The work of VRP, BSP, DGR, and AS is supported by the U.S. National Science
Foundation via grant PHY89-21025,
that of SCP by the U.S. Department of Energy, Nuclear Physics
Division, under contract W-31-109-ENG-38,
and that of JC by the U.S. Department of Energy.

\newpage
\begin{table}[t]
\caption{Comparison of microscopic and Skyrme-model energies.}
\begin{tabular}{lddddd}
 & $^8$n$(0^+)$ & $^7$n SOS & Magic  & $^{15}$N SOS$^{\rm a}$ &  W$_{\ell s}$\\
 &              &           & Nuclei &              &                 \\
 & MeV           & MeV       & $|\Delta E/E|_{rms}\%$ & MeV & MeV$^5$\\
\hline \\
GFMC             & -37.6(3) & 1.1(3) &    &   \\
CVMC             & -35.5(1) & 1.4(1) & --   & 6.1$^{\rm b}$  \\
\\
SkM                 & -47.4 & 3.0 &  1.1 & 6.3 & 130. \\
FPS-21              & -42.2 & 3.0 &  1.1 & 6.7 & 110. \\
Skyrme 1$^{\prime}$ & -38.7 & 2.9 &  1.8 & 6.9 & 120. \\
FPS                 & -32.5 & 3.5 &  1.2 & 6.7 & 110. \\
\\
\end{tabular}
\tablenotetext[1]{Experimental value 6.9 MeV deduced in Ref.\ \cite{pieper93}.}
\tablenotetext[2]{With Argonne v$_{14}$, Ref.\ \cite{pieper93}.}
\end{table}
\begin{figure}
\caption{
The transient energy $E(\tau)$ for the GFMC calculations of $^8$n$(0^+ )$ and
$^7$n(1/2$^-$ and 3/2$^-$)
as a function of the imaginary time $\tau$.}
\end{figure}

\begin{figure}
\caption{
Neutron density distribution for $^8$n, according to methods described in the
text.  The curves labeled A and B come from modified versions of the effective
interaction FPS21.}
\end{figure}

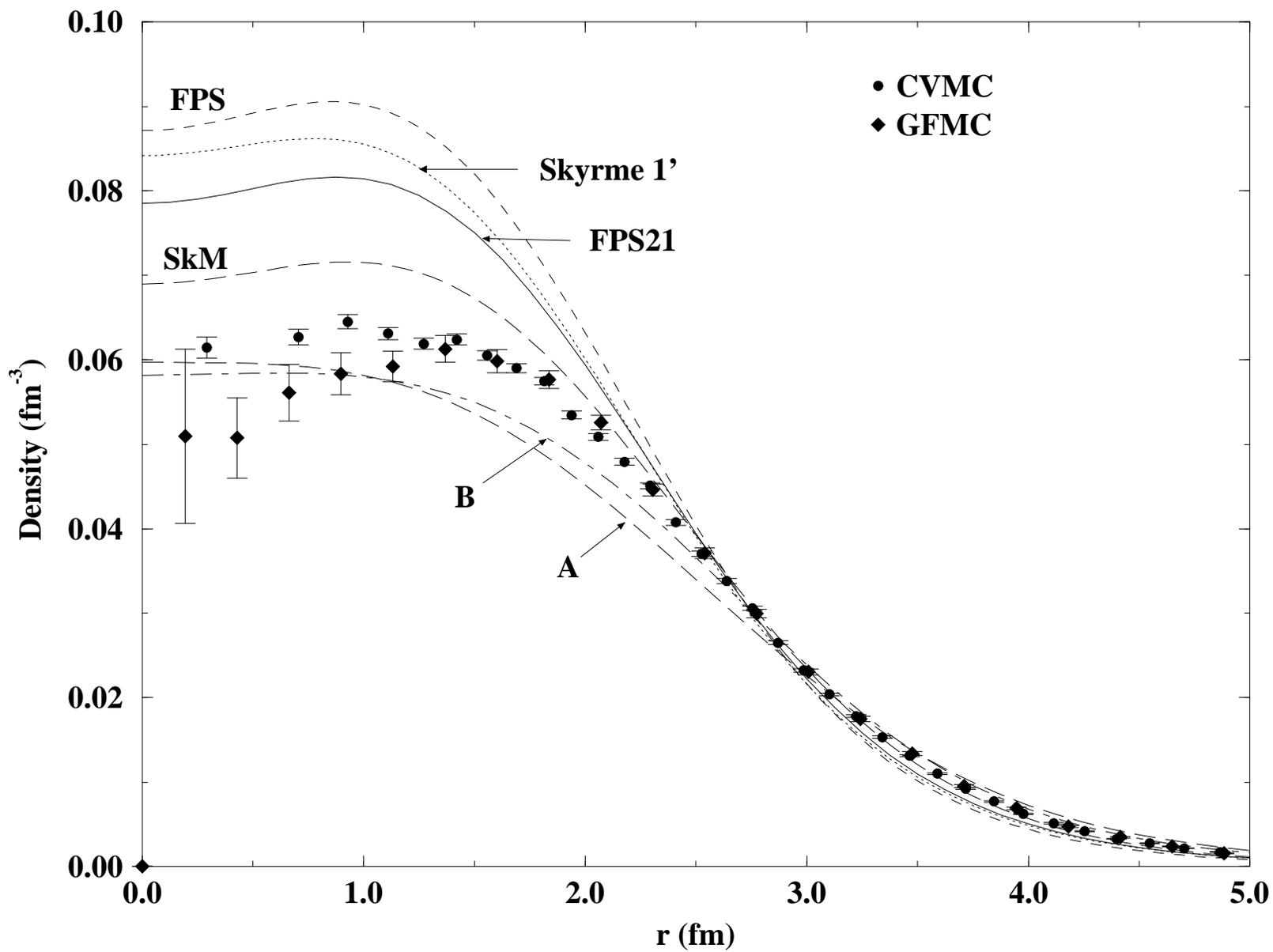
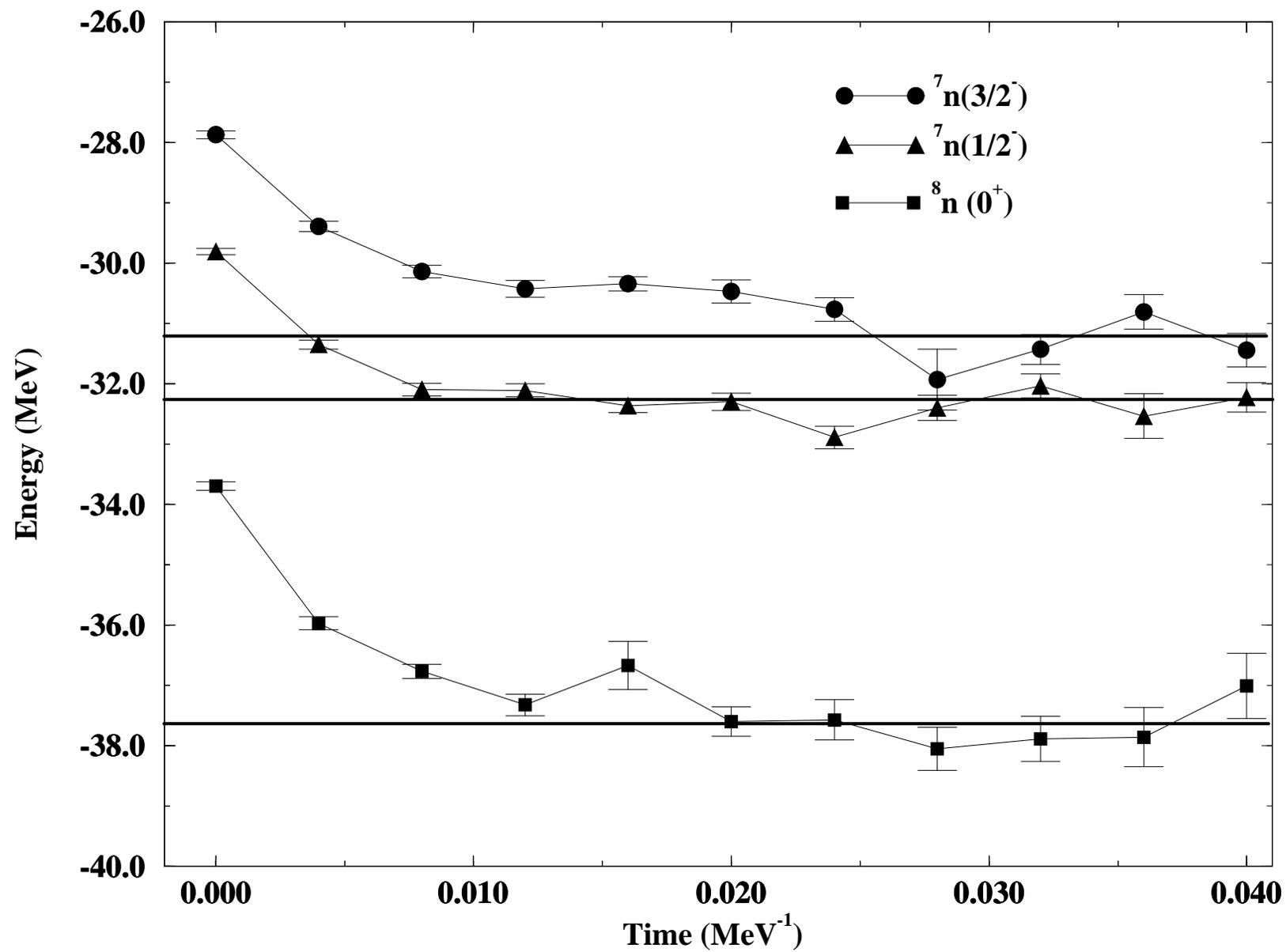
\end{document}